\documentclass[12pt]{article}
\usepackage{cite,epsfig,feynarts,amssymb,amsmath}

\newcommand{\Eq}[1]{Eq.\,(\ref{#1})}
\newcommand{\Eqs}[1]{Eqs.\,(\ref{#1})}
\newcommand{\Eqsand}[2]{Eqs.\,(\ref{#1}) and (\ref{#2})}

\newcommand{\Ref}[1]{Ref. \cite{#1}}
\newcommand{\Refs}[2]{Refs. \cite{#1, #2}}

\newcommand{\pref}[1]{(\ref{#1})}

\newcommand{\beq}{\begin{equation}}                
\newcommand{\eeq}{\end{equation}}        
\newcommand{\bea}{\begin{eqnarray}}               
\newcommand{\eea}{\end{eqnarray}}        
\newcommand{\bdm}{\begin{displaymath}}                 
\newcommand{\edm}{\end{displaymath}}                      
  
\newcommand{\non}{\nonumber}
\newcommand{\equal}[1]{\hspace{#1} & = & \hspace{#1}}

\newcommand{\smallH}{{\scriptscriptstyle H}}
\newcommand{\smallW}{{\scriptscriptstyle W}}        
\newcommand{\smallZ}{{\scriptscriptstyle Z}}    
        
\newcommand{\smallR}{{\scriptscriptstyle R}}

\newcommand{\f}{\frac}       
\newcommand{\fsl}{\hspace{-0.75em} / \hspace{0.25em}}     
 
\newcommand{\order}{{O}}

\newcommand{\mb}{m_b}        
        
\newcommand{\MT}{M_t}

\newcommand{\mw}{M_\smallW}
\newcommand{\MW}{M_\smallW}
\newcommand{\mz}{M_\smallZ}     
\newcommand{\mh}{M_\smallH}
     
\newcommand{\muw}{\mu_\smallW}  
\newcommand{\mut}{\mu_t} 
\newcommand{\mtbar}{\overline{M}_t}

\newcommand{\gev}{\, {\rm GeV}}

\newcommand{\MSbar}{ \overline{\rm MS} }

\newcommand{\gs }{ {g_s} }
\newcommand{\as }{ {\alpha_s} }         
\newcommand{\aem}{ {\alpha} }

\newcommand{\ew}{electroweak~}

\newcommand{\SWS}{ {s_\smallW^2} }      
\newcommand{\SWQ}{ {s_\smallW^4} } 
\newcommand{\CWS}{ {c_\smallW^2} }

\begin{document}

\thispagestyle{empty}
\rightline{CERN-TH/2001-213}
\rightline{TUM-HEP-426/01}
\rightline{MPI-PHT/2001-26}
\rightline{\today}
\vspace*{1.2truecm}
\bigskip

\centerline{\LARGE\bf  Complete electroweak  matching }
\vspace{.2cm}\centerline{\LARGE\bf for radiative B decays}
\vskip1truecm
\centerline{\large\bf Paolo Gambino$^a$ and Ulrich Haisch$^{b,c}$}
\bigskip
\begin{center}{
{\em $^a$ CERN, Theory Division, CH--1211 Geneve 23, Switzerland. }\\
\vspace{.3cm}
{\em $^b$ Max-Planck-Institut f\"ur Physik
  (Werner-Heisenberg-Institut),\\ 
F\"ohringer Ring 6, 80805 M\"unchen, Germany}\\
\vspace{.3cm}
{\em $^c$ Technische Universit\"at M\"unchen, Physik Dept.,\\
James-Franck-Str., D-85748 Garching, Germany}
}\end{center}
\vspace{1.5cm}

\centerline{\bf Abstract}
\vspace{1.cm}
We compute  the  complete two--loop $\order(\aem)$ Wilson coefficients
relevant for  radiative decays of the $B$ meson in the SM. This is a
necessary step in the calculation of the $\order(\aem\alpha_s^n \ln^n
\mb/\mw)$ corrections and improves on our previous analysis of \ew
effects in $B\to X_s \gamma$. We describe in detail several
interesting technical aspects of the calculation and include all
dominant QED matrix elements. In our final result, we neglect only
terms originated from the unknown $\order(\aem \as)$ evolution of the
Wilson coefficients and some suppressed two--loop matrix elements.
Due to the compensation among different effects, we find that
non--trivial \ew corrections decrease the branching ratio by about
3.6\% for a light Higgs boson, very close to our previous result. 
The corresponding up--to--date SM prediction for the branching ratio
with $E_\gamma>1.6 \gev$ is $(3.61\pm 0.30)\times 10^{-4}$.  
\vspace*{2.0cm}

\newpage
\section{Introduction}
The Standard Model (SM) calculation of the  branching ratio 
for the  inclusive radiative decay $B\to X_s \gamma$ --- BR$_\gamma$
in the following --- has reached a high degree of sophistication (see
\cite{gambino-misiak} for a complete list of references and an
up--to--date analysis). Besides  Leading Logarithmic $\order(\alpha_s^n
L^n)$ ($L=\ln m_b/\mw$) and Next--to--Leading Logarithmic (NLO) 
$\order(\alpha_s^n L^{n-1})$ QCD corrections and non--perturbative Heavy
Quark Effective Theory contributions, \ew effects are known to play a
non--negligible role \cite{previous,marciano,strumia,kagan,misiakQED}. 
In a previous work \cite{previous}, we have considered in detail the 
electroweak corrections to this process, devoting special attention to
the interplay between QCD and \ew effects. 
Photonic interactions generate logarithmically enhanced contributions 
which are suppressed by a factor $\aem/\as$ with respect to the
QCD ones. The leading QED effects are therefore
$\order(\aem\alpha_s^{n-1}L^n)$ and are known completely
\cite{misiakQED}, while genuine \ew corrections involving $Z^0$ and
$W$ bosons start at the next order in the resummed logarithmic
expansion. Ideally, one would like to have all these $\order(\aem
\alpha_s^n L^n)$ corrections under control. Since the $\order(\aem)$
contributions to the coefficients of the four quark operators are all
known \cite{lectures,previous}, this would entail the following steps
\cite{NNLO,previous}: 
\begin{itemize}
\item [(i)] the calculation of the two--loop $\order(\aem)$ matching
  conditions for the magnetic operators $Q_{7}^\gamma$ and $Q_8^g$ at
  a scale $\order(\mw)$; 

\item [(ii)] the QED--QCD evolution of the Wilson coefficients down to
  the $B$ mass scale, including the calculation of the two and
  three--loop $\order(\aem\as)$   anomalous dimension matrix; 

\item[(iii)] the calculation of the one--loop and two--loop 
QED matrix elements of  the various operators as well as of some 
yet unknown two--loop QCD matrix elements.
\end{itemize}

Our analysis in \cite{previous} was based on the simplifying
assumption that terms vanishing as
$s_\smallW\equiv\sin\theta_\smallW\to 0$ can be neglected, unless they
are enhanced by powers of the top mass $\MT$. In this case, introducing 
the ${\rm SU(2)_L}$ coupling $g$ and $\aem_\smallW=g^2/4\pi$, all \ew
corrections are in fact $\order(\aem_\smallW\alpha_s^n L^n)$ or
$\order(\aem \MT^2/\mw^2 \alpha_s^n L^n)$ and are included  by step
(i) only.  Moreover, the calculation of the Wilson coefficients
simplifies considerably. Although reasonable, this assumption should
be verified --- keep in mind that $s_\smallW^2\approx 0.23$. 
In particular, $Z^0$ boson corrections to the one--loop
$b\to s \gamma$ magnetic penguin diagrams give rise to
$\order(s_\smallW^2)$ terms  which are not formally suppressed by an
electric charge factor $Q_d^2=1/9$ or $Q_u |Q_d|=2/9$, unlike the purely
QED corrections of steps (ii) and (iii).  This happens, for instance,
because of the mass difference between $Z^0$ and $W$ bosons. Such
$\order(s_\smallW^2)$ terms originate at the \ew  scale and
affect only step (i).  

In this note we extend our calculation \cite{previous}  
and compute the full $\order(\aem)$ contribution to the Wilson
coefficients of the $b\to s $ magnetic operators, thus completing step
(i). The main difference (and technical hurdle) with respect to
\cite{previous} is due to the presence of virtual photons in the
two--loop SM diagrams. The resulting infrared (IR) divergences are
removed  in the matching with the effective low--energy theory
of quarks, photons and gluons. Several subtleties arise in the
calculation, mostly linked to the presence of unphysical
operators. This is explained in detail in Section 2, while  Section 3
deals with the QED--QCD evolution of the coefficients  and illustrates
how $\order(\aem \alpha_s^n L^n)$ effects should be taken into account
in the calculation of BR$_\gamma$. We also include all dominant
$\order(\aem)$ matrix elements and conclude reconsidering the SM
prediction of BR$_\gamma$.

\section{The $\order(\aem)$ matching} 
Let us briefly  recall the formalism. We work in the framework of an
effective low-energy theory with five active quarks, photons and
gluons, obtained by integrating out heavy degrees of freedom
characterized by a mass scale $M \ge \MW$. In the leading order of the
operator product expansion the effective off--shell Hamiltonian
relevant for the $b \to s \gamma$ and $b \to s g$ transition at a
scale $\mu$ is given by
\beq \label{hamiltonian}
{\cal H}_{\rm eff} = -\f{G_F}{\sqrt{2}} V_{ts}^\ast V_{tb} \left [ \;
  \sum_{i = 1}^{16} C_i (\mu) Q_i + C_7^\gamma (\mu) Q_7^\gamma + 
  C_8^g (\mu) Q_8^g \; \right ] \, .  
\eeq
Here $V_{ij}$ are the CKM matrix elements and $C_i (\mu)$, $C_7^\gamma
(\mu)$ and $C_8^g (\mu)$ denote the Wilson coefficients of the
following set of  gauge invariant operators \cite{Grinstein:1988pu,
Ciuchini:1998xe, misiakQED, Bobeth:1999mk} 
\bea \label{operatorbasis}
\begin{split}
& 
\begin{aligned}
Q_1 & = (\bar{s}_L \gamma_\mu T^a c_L) (\bar{c}_L \gamma^\mu T^a
b_L) \, , \\
Q_3 & = (\bar{s}_L \gamma_\mu b_L) \! \sum\nolimits_q \! (\bar{q}
\gamma^\mu q) \, , \\ 
Q_5 & = (\bar{s}_L \gamma_\mu \gamma_\nu \gamma_\rho b_L)
\! \sum\nolimits_q \!  (\bar{q} \gamma^\mu \gamma^\nu \gamma^\rho q)
\, , \\
Q_7 & = (\bar{s}_L \gamma_\mu b_L) \! \sum\nolimits_q \! Q_q (\bar{q}
\gamma^\mu q) \, , \\
Q_9 & = (\bar{s}_L \gamma_\mu \gamma_\nu \gamma_\rho b_L)
\! \sum\nolimits_q \! Q_q (\bar{q} \gamma^\mu \gamma^\nu \gamma^\rho
q) \, , \\
Q_7^\gamma & = \f{e}{16 \pi^2} \mb (\bar{s}_L \sigma^{\mu \nu} b_R)
F_{\mu \nu} \, , \\
Q_{11} & = \f{1}{e} \bar{s}_L \gamma^\mu b_L \partial^\nu F_{\mu \nu}
+ Q_7 \, ,
\end{aligned} \quad 
\begin{aligned}
Q_2 & = (\bar{s}_L \gamma_\mu c_L) (\bar{c}_L \gamma^\mu b_L) \, , \\
Q_4 & = (\bar{s}_L \gamma_\mu T^a b_L) \! \sum\nolimits_q \! (\bar{q}
\gamma^\mu T^a q) \, , \\
Q_6 & = (\bar{s}_L \gamma_\mu \gamma_\nu \gamma_\rho T^a b_L)
\! \sum\nolimits_q \!  (\bar{q} \gamma^\mu \gamma^\nu \gamma^\rho T^a
q) \, , \\ 
Q_8 & = (\bar{s}_L \gamma_\mu T^a b_L) \! \sum\nolimits_q \! Q_q (\bar{q}
\gamma^\mu T^a q) \, , \\
Q_{10} & = (\bar{s}_L \gamma_\mu \gamma_\nu \gamma_\rho T^a b_L)
\! \sum\nolimits_q \! Q_q (\bar{q} \gamma^\mu \gamma^\nu \gamma^\rho
T^a q) \, , \\
Q_8^g & = \f{\gs}{16 \pi^2} \mb (\bar{s}_L \sigma^{\mu \nu} T^a b_R)
G_{\mu \nu}^a \, , \\
Q_{12} & = \f{1}{\gs} \bar{s}_L \gamma^\mu T^a b_L D^\nu G_{\mu
\nu}^a + Q_4 \, ,
\end{aligned} \\ 
& 
\begin{aligned}
Q_{13} & = \f{i e}{16 \pi^2} \Big [ \bar{s}_L \stackrel{\leftarrow}{D
  \fsl} \sigma^{\mu \nu} b_L F_{\mu \nu} - F_{\mu \nu} \bar{s}_L
\sigma^{\mu \nu} D \fsl b_L \Big ] + Q_7^\gamma \, , \\
Q_{14} & = \f{i \gs}{16 \pi^2} \Big [ \bar{s}_L
\stackrel{\leftarrow}{D \fsl} \sigma^{\mu \nu} T^a b_L G_{\mu \nu}^a -
G_{\mu \nu}^a \bar{s}_L T^a \sigma^{\mu \nu} D \fsl b_L \Big ] + Q_8^g
\, , 
\end{aligned} \quad 
\begin{aligned}
Q_{15} & = \f{1}{16 \pi^2} \mb \bar{s}_L D \fsl D \fsl b_R \, , \\
Q_{16} & = \f{i}{16 \pi^2} \bar{s}_L D \fsl D \fsl D \fsl b_L \, ,
\end{aligned}
\end{split}
\eea
where $e$ ($\gs$) is the electromagnetic (strong) coupling constant,
$q_{L, R}$ are the chiral quark fields, $F_{\mu \nu}$ ($G_{\mu
\nu}^a$) is the electromagnetic (gluonic) field strength tensor,
$D_\mu$ is the covariant derivative of the gauge group ${\rm SU(3)_C
\! \times \! U(1)_Q}$ and $T^a$ are the colour matrices, normalized
so that $\mbox{Tr} (T^a T^b) = \delta^{ab}/2$. The $s$--quark mass
is neglected in \Eq{operatorbasis} and in the following. Notice that
at the order we are going to work it is not necessary to consider the
analogues of $Q_1$ and $Q_2$ involving the $u$--quark instead of the
$c$--quark. 
  
The above set of operators  closes off--shell under  QCD and QED
renormalization, up to non-physical (evanescent) operators that vanish
in four dimensions \cite{Grinstein:1988pu,Bobeth:1999mk}. It consists
of the current-current operators $Q_1$--$Q_2$, the QCD penguin
operators $Q_3$--$Q_6$, the electroweak penguin operators
$Q_7$--$Q_{10}$ and the magnetic moment type operators $Q_7^\gamma$
and $Q_8^g$. It is the QED renormalization that forces us to introduce
the operators $Q_7$--$Q_{10}$, in which the sum of the quark flavors
is weighted by the electric charges $Q_q$. The remaining six operators
$Q_{11}$--$Q_{16}$, characteristic of the process $b \to s \gamma \,
(g)$, were chosen in such a way that they vanish on--shell up to total
derivatives.  
Only operators of dimension five or six are retained. Higher dimension
operators are suppressed by at least one power of $\mb^2/\MW^2$, while
those of lower dimensionality can be removed by choosing suitable
renormalization conditions in the full theory
\cite{Altarelli:1974ex}. In the present case this is achieved by
requiring that all flavor off--diagonal quark two--point functions
which appear at the one--loop level in the full theory vanish when the
equations of motion (EOM) are applied, i.e.\ by using LSZ on--shell
conditions on the external quark lines. We also use these
renormalization conditions for internal virtual quarks in the full SM
calculation and implement in this way  a gauge invariant
$\order(\alpha)$ definition of the CKM matrix \cite{CKM}. 

In order to obtain  the Wilson coefficients of the magnetic  operator
$Q_7^\gamma$, we calculate the off--shell amplitude $b \to s \gamma$ 
in the full SM and in the effective theory at $\order(\alpha)$ and
match the two results. 
Retaining only the leading terms in $1/\MW^2$ the off--shell amplitude
in the full theory can be written in the following form\footnote{In
\Eqsand{AfullQi}{AeffQi}, the sum runs over $Q_1$--$Q_{16}$,  
$Q_7^\gamma$, $Q_8^g$ and, as we will explain later on, some
evanescent operators.}
\beq \label{AfullQi}
{\cal A}_{{\rm full}} = -\f{G_F}{\sqrt {2}} V_{ts}^\ast V_{tb} \sum_{i}
A_i \left \langle s \gamma \left | Q_i \right | b \right \rangle^{(0)}
\, ,
\eeq
where $\left \langle s \gamma \left | Q_i \right | b \right
\rangle^{(0)}$ are the tree--level matrix elements of the operators in
\Eq{operatorbasis}. The perturbative expansion of the coefficients
$A_i$ reads
\beq \label{Aexpansion}
A_i  = A_i^{(0)}  + \f{\aem}{4 \pi} A_{i, e}^{(1)}  \, .
\eeq
We calculate analytically the relevant one and two--loop amplitudes 
starting from the diagrams generated by {\it FeynArts 2.2}
\cite{feynarts} and retaining only  terms 
which project on  $Q_7^\gamma$ after use of the EOM.
All ultraviolet (UV) divergences in $A_{7, e}^{\gamma (1)}$ are removed by
\ew renormalization. We follow closely the procedure outlined in
\cite{previous,bbbar}. The only additional ingredients not explicitly
given in those papers are
the right--handed down quark wave function renormalization
($c_\smallW^2=1-s_\smallW^2$) 
\beq
\delta (Z_d^\smallR)_{ij}=
{g^2 \over 16 \pi^2} \frac{Q_d^2 s_\smallW^4}{c_\smallW^2} \delta_{ij}
\left[\frac1{\epsilon}- \frac12  -\ln \frac{\mz^2}{\mu^2}\right] \, , 
\label{dzr}
\eeq
and  the complete  $b$--quark on--shell mass counterterm
($x_t=\mtbar^2/\mw^2$)
\beq \label{dmb} 
\begin{split}
\f{\delta \mb}{\mb} & = \f{g^2}{16 \pi^2}  \left
  (\f{\bar{\mu}^2}{\mtbar^2} \right )^\epsilon \left[ 
\f{1}{8 \epsilon} (3 x_t - 2) -\f{2 + 9 x_t - 5 x_t^2}{16 (x_t - 1)} -
\f{2 - 7 x_t + 2 x_t^2}{8 (x_t - 1)^2} \ln x_t \right. \\ 
& \hspace{1cm} \left. + \f{1}{\CWS} \left ( \f{1}{16} -\f{5}{12} \SWS +
\f{5}{18} \SWQ + \left ( \f{1}{8} +\f{\SWS}{2} - \f{\SWQ}{3} \right )
\left (\ln \f{\mz^2}{\mtbar^2} - \f{1}{\epsilon} \right ) \right )
\right ] \, , 
\end{split}
\eeq
which had not been given in \cite{previous,bbbar}.
 Notice that 
\Eqsand{dzr}{dmb} do {\it not} include the photon contribution, for a reason
that will become clear in a moment. The top mass $\mtbar$ is renormalized
on--shell as far as \ew effects are concerned, while we use an
$\overline{\rm MS}$ definition at a scale $\mu$ for the QCD renormalization. 

We work in the background field gauge (BFG) \cite{bfg}. This reduces
the number of diagrams to be considered. Moreover, if the electric
charge is normalized at $q^2=0$, as it is natural to do
\cite{marciano}, its counterterm  cancels identically against the
background photon wave function renormalization factor, due to the 
BFG Ward identity \cite{denner}. The same holds in the case of
an external gluon in the $\MSbar$ scheme. The regularization  problems
related to the definition of $\gamma_5$ in $n=4-2\epsilon$ dimensions
are avoided as described in \cite{previous} and we employ the naive
dimensional regularization scheme with anticommuting $\gamma_5$ (NDR)
throughout the paper. 

For what concerns the regularization of the IR divergences, we have
adopted two different methods  and found identical results for 
the Wilson coefficients. In the first method the IR divergences are
regulated by  quark masses (see \cite{Ciuchini:1998xe}), while the
second method consists in  using dimensional regularization for both
UV and IR divergences \cite{Bobeth:1999mk}. 

\begin{figure}[t]
\begin{center}
\vspace{-1cm}
\input{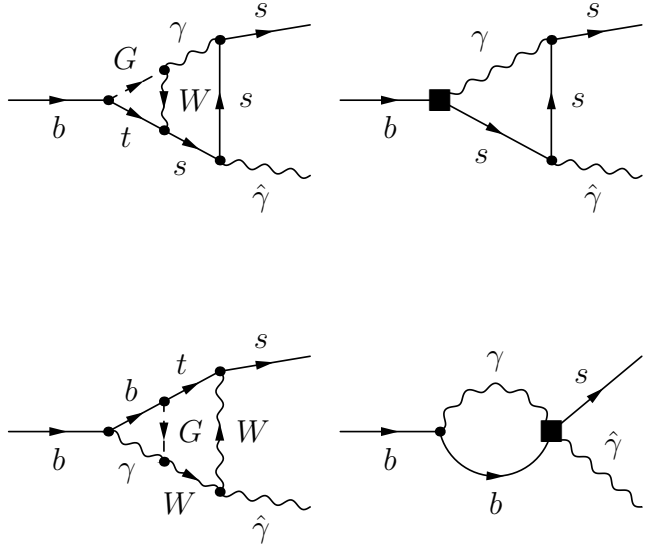}
\end{center}
\vspace{-1cm}
\caption{\sf Subdiagrams involving the coupling of quantum photons
  $\gamma$ to the $W$ and its corresponding Goldstone boson (left hand
side) contribute to gauge variant operators (right hand side), as
explained in the text. $\hat{\gamma}$ denotes a background
photon.}
\label{bsA}
\end{figure}
A  second step involves the calculation of the off--shell amplitude in the
QED effective theory. In general, we need effective vertices
with both background and quantum photons. Interestingly,
the latter introduce some gauge variant operators at $\order(\aem)$.
In fact, on the full theory side there are heavy particle subdiagrams 
(see Fig.~\ref{bsA}) that are coupled to quantum photons and
contribute  to gauge variant operators not included in the operator
basis of \Eq{operatorbasis}. This is due to the $R_\xi$ gauge coupling of 
quantum photons with $W$ and Goldstone bosons and is different from
what happens in the case of the off--shell $\order(\as)$ matching 
\cite{Ciuchini:1998xe,Bobeth:1999mk}. Indeed, at $\order(\as)$
only quark--gluon couplings and trilinear quantum--quantum--background
gluon couplings are relevant and no gauge variant operator is induced.
The appearance of gauge variant operators in the SM amplitudes is not
surprising \cite{collins} (see  \cite{misiak-munz} for an example). 

We have explicitly verified that it is not necessary to take any 
gauge variant operator into account on the effective theory
side. This
follows from well--known theorems on the renormalization of gauge
invariant operators\footnote{The theorems apply to Yang--Mills
theories, but extend to the full SM after imposing the anti--ghost
equation \cite{antighost}.} \cite{collins,barnich}:  gauge variant  
operators that mix with gauge invariant operators can be chosen so
that they are all BRST--exact, i.e.\ they can be written as the
BRST--variation of some other operators, modulo terms vanishing by 
the EOM. Therefore, while gauge invariant operators generally mix  into
gauge variant operators\footnote{At $\order(\aem)$ 
the operators in our basis do {\it not} actually mix into gauge
variant operators.}, the opposite is not true. Since we are eventually
interested in the matrix elements of physical operators only, we do
not need to include  gauge variant effective operators in our basis.  
Of course this holds only as long as the regularization respects the
symmetries, like it is in our case. 

In a similar way and because of the same theorems, the operators that
vanish by EOM in the basis (\ref{operatorbasis}) do not mix into the
physical operators of the same basis, and the renormalization mixing
matrix is block triangular. This property drastically simplifies the
computation at hand, as we will see in a moment. In particular, the
renormalization mixing matrix $\hat{Z}$ is such that $Z_{ij} = 0$ when
$Q_i$ is EOM--vanishing and $Q_j$ is a physical operator.

We have seen that effective vertices involving quantum photons are
induced in the calculation. Even though contributions
to gauge variant operators turn out to be irrelevant, the distinction is 
important in the case that the calculation is performed using quark masses
to regularize IR divergences. For example, it turns out that the gauge
invariant part of the off--shell $b \to s \gamma$ effective vertex
depends on whether the external photon is quantum  or background. This
can be explained by noting that  the
operators involving only background fields are combinations of truly
gauge invariant operators and of  operators containing also quantum
gauge fields.  This follows, e.g. from a decomposition of the kind
$\hat{D}\fsl =D \fsl + i e Q_q Q \fsl$, where we used a hat to denote
covariant background derivative,  $Q_\mu$ for the quantum photon
field, and $Q_q$ for the electric charge. The operators containing
quantum gauge bosons eventually decouple from the calculation, as they
are not gauge invariant. Because of the above decomposition,  the
coefficients of the operators involving only background fields are
related to the coefficients of the operators in (\ref{operatorbasis}),
as can be seen using Slavnov--Taylor identities of the kind used in
\cite{algren}.  

The effective theory calculation depends crucially on the IR
regularization. The first method mentioned above (quark masses as
regulator) can be applied at the diagrammatic level
\cite{Ciuchini:1998xe}. The effective theory diagrams are obtained by
replacing hard (heavy mass) subdiagrams in the two--loop SM amplitudes
with their Taylor expansions with respect to their external
momenta. In principle, this method does not require a discussion of
the effective operators. On the other hand,  it is relatively
complicated to implement.  Here, we limit our discussion to  the
second method only, following \cite{Bobeth:1999mk}. In order to get
the renormalized off--shell amplitude on the effective theory side, we
need to reexpress \Eq{hamiltonian} in terms of renormalized
quantities. The relations between the bare and the QED renormalized
quantities are as follows 
\beq
e_0 = Z_e e \, , ~~ m_{b, 0} = Z_{\mb} \mb \, , ~~ A_0^\mu = Z_e^{-1}
A^\mu \, , ~~ q_0 = Z_q^{1/2} q \, , ~~ C_{i, 0} = \sum_j C_j Z_{ji}
\, ,
\eeq
with $Z_e$, $Z_{\mb}$, $Z_q$ and $Z_{ij}$ the renormalization constant
of the charge, the $b$--quark mass, the quark fields and the Wilson
coefficients, respectively. The relation between the renormalization
of the gauge field and of the electric charge is a direct consequence
of the QED Ward identity.   

After renormalization the off--shell amplitude in the effective theory
is given by
\beq \label{AeffQi}
{\cal A}_{{\rm eff}} = -\f{G_F}{\sqrt {2}} V_{ts}^\ast V_{tb} \sum_{i,j
} C_j Z_{ji} \widetilde{Z}_i
\left \langle s \gamma
\left | Q_i \right | b \right \rangle \, ,
\eeq
where $\widetilde{Z}_i
$ denotes a product of $Z_e$, $Z_m$ and $Z_q$ depending on the particular
structure of the operator $Q_i$ and the Wilson coefficients may be
expanded in powers of $\aem$ as follows 
\beq \label{Cexpansion}
C_i (\mu ) = C_i^{(0)} (\mu) + \f{\aem}{4 \pi} C_{i, e}^{(1)} (\mu) \, .
\eeq

As long as we are only interested in the Wilson coefficient of the
magnetic photon penguin operator, it is sufficient to keep only 
terms proportional to $\left \langle s \gamma \left | Q_7^\gamma
\right | b \right \rangle$ in \Eq{AeffQi}. Using the short hand
notation $\left \langle Q_7^\gamma \right \rangle \equiv \left \langle
s \gamma   \left | Q_7^\gamma \right | b \right \rangle$, the part of
the off--shell amplitude in the effective theory needed for the
matching of $C_7^\gamma$ is then written as
\beq \label{AeffQ7g}
{\cal A}_{{\rm eff}} \sim -\f{G_F}{\sqrt {2}} V_{ts}^\ast V_{tb} \left
[ Z_q Z_{\mb} \sum_j C_j Z_{j, 7 \gamma} + Z_q (Z_{\mb} - 1) \sum_j
C_j Z_{j, 13} \right ] \left \langle Q_7^\gamma \right \rangle \, ,
\eeq
where the second term proportional to $Z_{j, 13}$ originates from the
renormalization of the operator $Q_{13}$.

Notice that the QED quark field renormalization on the effective side
can be avoided as it cancels in the matching against the photon
contribution to the corresponding term  in the SM. The same applies to
the renormalization of the $b$--quark mass, which is retained only up
to linear terms. Consequently, after checking the cancellation of
the UV divergences in (\ref{Aexpansion}) we have omitted the photon
contributions in \Eqsand{dzr}{dmb} and simultaneously set $Z_{\mb}$
and $Z_q$ to unity in the effective theory. This simplifies 
the following considerations.

Adopting the $\MSbar$ scheme for the operator renormalization the
corresponding renormalization constants can be written as
\beq\label{zeta}
Z_{ij} = \delta_{ij}  + \f{\aem}{4 \pi}
\f{1}{\epsilon} Z_{ij}^{(1)} + Z_{i j}^{(0)} \, . 
\eeq
The renormalization constants $Z_{ij}^{(1)}$ are found by calculating the
UV divergent parts of Feynman diagrams in the effective theory. Within
the scope of this computation, it is essential to carefully
distinguish UV from IR singularities. As explained in
\Ref{Chetyrkin:1998fm}, this can be done most easily by introducing a
common mass parameter into all the propagator denominators including
the photon ones. All renormalization constants in the effective theory
up to two loops are known from previous anomalous dimension
calculations \cite{kagan, misiakQED}. As we shall see later on, only
five entries of the anomalous dimension matrix are relevant in the
present computation and we have recalculated these elements to
check the results mentioned above. Our results are in full agreement
with \Refs{kagan}{misiakQED} and we will give the numerical values of
the required renormalization constants below
\Eqsand{NLOmatchingEqC7g}{NLOmatchingEqC8g}. 

The last term in \Eq{zeta} implies a finite renormalization of $Q_i$
at zeroth order. Indeed, in situations where evanescent operators are
present, the standard practice is to extend the $\MSbar$ scheme and to allow for a finite
operator renormalization. The finite terms $Z_{i j}^{(0)}$ differ from
zero when  $Q_i$ is an evanescent operator and $Q_j$ is not, and their
values are fixed by requiring that renormalized matrix elements of
evanescent operators vanish in $n = 4$ dimensions
\cite{Buras:1990xd,evmisiak}.  This requirement also ensures that
evanescent operators do not mix into physical ones
\cite{Buras:1990xd}.  
Furthermore, in the case  of the $b\to s \gamma$ calculation, it is
well--known \cite{Ciuchini:1993ks} that some four quark operators can
mix into the magnetic operators through one--loop diagrams at zeroth
order in $\aem$ and $\as$. Thus, not only we have finite terms in
\Eq{zeta}, but they appear at the lowest order in the coupling
constant.  

The computation of the  necessary matrix elements on the effective
side is trivial, as we can set all the light particles masses to
zero\footnote{ We include only terms that are linear in the $b$--quark
  mass. They originate from use of  the EOM only.}. 
Accordingly, all loop diagrams on the effective
side vanish in dimensional regularization, because of the cancellation
between UV and IR divergences. 
Therefore only the tree--level matrix
elements $\left \langle Q_i \right \rangle^{(0)}$ are different from
zero and higher order matrix elements do not play any role in the
matching. Notice that due to the cancellation of UV and IR
singularities the UV counterterms present in the tree--level matrix
elements reproduce precisely the IR divergences in the effective
theory. Furthermore, the IR divergence on the effective side has to be
equal to the IR singularity on the SM side, to guarantee that the
final results of the Wilson coefficients are free of IR
poles. Eventually, all $1/\epsilon$ poles cancel out in $C_7^\gamma$,
if the full and the effective theory are matched in the correct way.

Bearing all this in mind, we are now able to extract from \Eq{AeffQ7g}
those terms which are actually needed to calculate the $\order (\aem)$
correction to the Wilson coefficient of the magnetic operator. First
of all, we have to perform the tree--level matching by computing the
relevant diagrams for the various operator insertions. Only $C_2$,
$C_7^\gamma$, $C_8^g$ and $C_{11}$--$C_{16}$ are found to be
non--vanishing at leading order. However, due to the triangularity 
of the mixing matrix the coefficients $C_{11}$--$C_{16}$ do not
contribute to the first term in \Eq{AeffQ7g} and therefore will not
affect the matching conditions at the next order. Furthermore, as
we set $Z_{\mb}$ equal to one also the term proportional to $(Z_{\mb}
- 1)$ in \Eq{AeffQ7g} does not contribute to $C_7^\gamma$ at $\order
(\aem)$. Using $Z_q =1$ we thus  obtain
\beq \label{AeffC7gO(a)}
\begin{split}
{\cal A}_{{\rm eff}} \sim -\f{G_F}{\sqrt {2}} V_{ts}^\ast V_{tb}
\Bigg ( C_7^{\gamma (0)}  +  \f{\aem}{4 \pi} \Bigg [ &C_{7,
    e}^{\gamma (1)} +  \f{1}{\epsilon} \left ( Z_{2, 7
      \gamma}^{(1)
} C_2^{(0)} + Z_{7\gamma, 7 \gamma}^{(1)
} C_7^{\gamma (0)} \right ) \\ 
&+ \sum_i Z_{E_i, 7 \gamma}^{(0)}  C_{i, e}^{E, (1)}  \Bigg ] \Bigg )
\left \langle Q_7^\gamma \right \rangle^{(0)} \, .
\end{split}
\eeq
It is quite remarkable that, with the exception of the last term, only
{\it physical} operators play a role in this expression, even though
the calculation has been performed off--shell. 

The matching procedure between the full and the effective theory
establishes the initial conditions for the Wilson coefficients a scale
$\muw = \order (\MW)$.  
Comparing \Eqs{AfullQi},
\pref{Aexpansion} and \pref{AeffC7gO(a)}, the matching condition ${\cal
A}_{\rm full} (\muw) = {\cal A}_{\rm eff} (\muw)$ translates into the
following  identities
\begin{align} 
\label{LOmatchingEqC7g}
C_7^{\gamma (0)} (\muw) &= A_7^{\gamma (0)} (\muw) \,
, \\  
\label{NLOmatchingEqC7g}
C_{7, e}^{\gamma (1)} (\muw) &= A_{7, e}^{\gamma (1)}
(\muw) - \f{1}{\epsilon} \left ( Z_{2, 7 \gamma}^{(1)
} C_2^{(0)} (\muw) + Z_{7 \gamma, 7 \gamma}^{(1)
} C_7^{\gamma (0)} (\muw) \right ) \non \\ 
& \hspace{0.5cm} -\sum_i  Z_{E_i, 7 \gamma}^{(0)} C_{i, e}^{E
  (1)}(\muw) \, ,  
\end{align}
from which the Wilson coefficient of the magnetic operator up to
$\order (\aem)$ can be calculated. The leading order initial 
condition for the Wilson coefficient of $Q_2$ is simply  $C_2^{(0)}
(\muw) = 1$ and the elements of the mixing matrix needed for the next
leading order matching of $C_7^\gamma$ are $Z_{2, 7 \gamma}^{(1)
}= -58/243$ and $Z_{7 \gamma, 7 \gamma}^{(1)
} = 8/9$. Note that the renormalization constant $Z_{2, 7 \gamma}^{(1)
}$, related to
the mixing of the operators $Q_2$ and $Q_7^\gamma$, is obtained from a
two--loop calculation, as opposed to $Z_{7 \gamma, 7 \gamma}^{(1)
}$ which only involves a one--loop calculation. Whereas $Z_{7 \gamma, 7
\gamma}^{(1)}$ is regularization and renormalization scheme
independent,  $Z_{2, 7 \gamma}^{(1)}$ is scheme dependent. 
The value for $Z_{2, 7 \gamma}^{(1)}$ given above corresponds to 
the NDR scheme  --- see \cite{lectures}. 
Notice also that in \Eq{NLOmatchingEqC7g} the 
$\order (\epsilon)$ terms of $C_7^{\gamma (0)}$
yield a finite contribution when  combined with the $1/\epsilon$
pole proportional to $Z_{7\gamma, 7 \gamma}^{(1)}$. 
Indeed, the leading order matching needs to be performed up to $\order
(\epsilon)$. Explicit formulas for the initial condition of
$C_7^{\gamma (0)}$ including $\order (\epsilon)$ terms can be found in
\cite{Greub:1997hf, Buras:1998xf}.

For what concerns the last term in \Eq{NLOmatchingEqC7g}, it is
necessary to introduce the following evanescent operators
\beq \label{evanescent}
\begin{split}
Q_1^E &= (\bar{s}_L \gamma_\mu b_L) \sum\nolimits_q (\bar{q}_L
\gamma^\mu q_L) + (1 + a_1 \epsilon) \left (\f{1}{3} Q_3 - \f{1}{12} Q_5
\right ) \, , \\ 
Q_2^E &=  (\bar{s}_L \gamma_\mu b_L) \sum\nolimits_q Q_q (\bar{q}_L
\gamma^\mu q_L) + (1 + a_2 \epsilon) \left (\f{1}{3} Q_7 - \f{1}{12} Q_9
\right ) \, , \\ 
Q_3^E &= (\bar{s}_L \gamma_\mu b_L) \sum\nolimits_q Q_q (\bar{q}_R
\gamma^\mu q_R) - (1 + a_3 \epsilon) \left (\f{4}{3} Q_7 - \f{1}{12} Q_9
\right ) \, ,  
\end{split}
\eeq
where $a_i$ are arbitrary constants. In NDR inserting these operators
into the one--loop $b\to s \gamma$ penguin diagrams yields $Z_{E_1, 7
  \gamma} = 4/9$, $Z_{E_2, 7 \gamma} = -4/27$ and $Z_{E_3, 7 \gamma} =
4/27$. Notice that the last term in \Eq{NLOmatchingEqC7g}  does not
depend on the special choice of evanescent operators adopted above,
but it does depend on the choice of physical operators. For instance,
in the operator basis of \cite{lectures}, all evanescent operators
that project on $Q_7^\gamma$ have vanishing Wilson coefficients, both
at $\order(\aem)$ and $\order(\as)$. Therefore, in this basis
evanescent operators do not affect the matching equations and is it
not necessary to introduce them in \Eq{zeta}. Curiously, in the
operators basis of \Eq{operatorbasis} the same holds only at
$\order(\as)$.    
 
We have verified that all $1/\epsilon$ poles cancel in
\Eq{NLOmatchingEqC7g}, and that the result for $C_{7, e}^{\gamma 
(1)}$ coincides  with the one obtained using quark masses for the
 IR regularization. In the latter case evanescent operators 
do not play any role in the matching, as their contribution to
the matrix elements cancels against a corresponding term stemming
from the finite renormalization $Z_{E_i, 7 \gamma}^{(0)}$. 

Let us now turn to the matching for the Wilson coefficient of the
chromomagnetic penguin operator $Q_8^g$. The calculation for the $b\to
s$ gluon off--shell amplitude proceeds in the same way as above. 
Adopting the notation $\left \langle Q_8^g \right \rangle
\equiv \left \langle s g \left | Q_8^g \right | b \right \rangle$,
we see that the analogue of \Eq{AeffC7gO(a)} is
\beq \label{AeffQ8gO(a)}
\begin{split}
{\cal A}_{{\rm eff}} \sim -\f{G_F}{\sqrt {2}} V_{ts}^\ast V_{tb} \Bigg
( C_8^{g (0)} + \f{\aem}{4 \pi} \Bigg [ &C_{8, e}^{g (1)} +
\f{1}{\epsilon} \left ( Z_{2, 8 g}^{(1)} C_2^{(0)} + Z_{7 \gamma,
8 g}^{(1)} C_7^{\gamma (0)} + Z_{8 g, 8 g}^{(1)} C_8^{g (0)}
\right)  \\
&+ \sum_i Z_{E_i, 8 g}^{(0)}  C_{i, e}^{E(1)}  \Bigg ] \Bigg )
\left \langle Q_8^g \right \rangle^{(0)} \, ,
\end{split}
\eeq
from which we obtain 
\bea
\label{LOmatchingEqC8g}
C_8^{g (0)} (\muw) \equal{-0.2cm} A_8^{g (0)} (\muw) \, , \\
\label{NLOmatchingEqC8g}
C_{8, e}^{g (1)} (\muw) \equal{-0.2cm} A_{8, e}^{g (1)} (\muw) -
\f{1}{\epsilon} \left ( Z_{2, 8 g}^{(1)} C_2^{(0)} (\muw) + Z_{7
    \gamma, 8 g}^{(1)} C_7^{\gamma (0)} (\muw) + Z_{8 g, 8 g}^{(1)}
  C_8^{g (0)} (\muw) \right) \non \\ 
&& \hspace{-0.2cm} - \sum_i  Z_{E_i, 8 g}^{(0)} C_{i, e}^{E(1)}(\muw)
\, ,   
\eea
where $Z_{2, 8 g}^{(1)} = -23/81$, $Z_{7 \gamma, 8 g}^{(1)} =
-4/3$ and $Z_{8 g, 8 g}^{(1)} = 4/9$. The renormalization constants
which describe the mixing of evanescent operators into physical ones
read  $Z_{E_1, 8 g} = -4/3$, $Z_{E_2, 8 g} = 4/9$ and $Z_{E_3, 8 g} =
-4/9$.  Again, all IR poles cancel in $C_{8, e}^{g (1)}$ and the
result coincides with the one obtained with the other method.  

We now recall that the relevant quantity entering the calculation of
BR$_\gamma$ is not $C_7^\gamma(\mu_b)$  with  $\mu_b=\order(m_b)$ but
a combination  $C_7^{\gamma, {\rm eff}}(\mu_b)$ of this Wilson
coefficient and of the coefficients of the four quark 
operators. This combination is the coefficient of 
$\langle Q_7^\gamma\rangle^{(0)}$ calculated on--shell.
It follows from this definition that,
unlike $C_7^\gamma$, the effective coefficient
is regularization scheme independent at LO \cite{Ciuchini:1993ks} 
and does not depend on the basis of physical operators.  In NDR    
the two  combinations relevant for $B\to X_s \gamma $ and $B\to X_s g$
are \cite{misiakQED} 
\beq
\begin{split}
C_{7}^{\gamma, {\rm eff}}(\mu) & = C_7^\gamma(\mu) + \sum_{i = 1}^{10}
y_i \,C_i(\mu) \, , \\
C_{8}^{g, {\rm eff}}(\mu) & = C_8^g(\mu) + \sum_{i = 1}^{10}
z_i \,C_i(\mu) \, , 
\end{split}
\label{cieff}
\eeq
where $y = (0, 0, -\f{1}{3}, -\f{4}{9}, -\f{20}{3}, -\f{80}{9},
\f{1}{9}, \f{4}{27}, \f{20}{9}, \f{80}{27})$ and $z = (0, 0, 1,
-\f{1}{6}, 20, -\f{10}{3}, -\f{1}{3}, \f{1}{18}, -\f{20}{3},
\f{10}{9})$. 
The $\order(\aem)$ contributions to the
Wilson coefficients at $\mu=\mw$ in our operator basis
\Eq{operatorbasis} can be found from those in the operator basis of
\cite{lectures} after a basis transformation which in four dimensions
is simply  
\beq \label{transf}
\vec{C}(\mu) = \hat{{\cal R}}^T \vec{C}^\prime(\mu) \, ,
\eeq
where $\vec{C}^\prime$  are the Wilson coefficients in the basis of
\cite{lectures}.  They are given in Eqs.~(8.111)--(8.117) of that
review. The matrix $\hat{\cal R}$ is the extension of the same matrix
of \cite{Chetyrkin:1998gb} and is needed only for the physical
operators $Q_1$--$Q_{10}$, $Q_7^\gamma$ and $Q_8^g$:
\beq
\hat{{\cal R}}=\left(
\begin{array}{rrrrrrrrrrrr}
2&\frac{1}{3}&0&0&0&0&0&0&0&0&0&0\\
0&1&0&0&0&0&0&0&0&0&0&0  \\
0&0&-\frac{1}{3} &0&\frac{1}{12}&0&0&0&0&0&0&0  \\
0&0&- \frac{1}{9} &- \frac{2}{3} &\frac{1}{36}&\frac{1}{6}&0&0&0&0&0&0  \\
0&0&\frac{4}{3}&0&- \frac{1}{12}  &0&0&0&0&0&0&0  \\
0&0&\frac{4}{9}&\frac{8}{3}&- \frac{1}{36}  &-\frac{1}{6} &0&0&0&0&0&0  \\
0&0&0&0&0&0&2&0&-\frac{1}{8} &0&0&0  \\
0&0&0&0&0&0&\frac{2}{3}&4&- \frac{1}{24}&- \frac{1}{4} &0&0  \\
0&0&0&0&0&0&- \frac{1}{2}  &0&\frac{1}{8}&0&0&0  \\
0&0&0&0&0&0&- \frac{1}{6}  &-1&\frac{1}{24}&\frac{1}{4}&0&0  \\  
0&0&0&0&0&0&0&0&0&0&1&0  \\  
0&0&0&0&0&0&0&0&0&0&0&1    
\end{array}\right) \, .
\eeq
In fact, beyond the leading order,  the operator basis must be
supplemented by a definition of the evanescent operators. This
definition corresponds to a choice of scheme and it is different, for
instance,  in the {\it standard} basis of \cite{lectures} and in the
operator basis of \cite{Chetyrkin:1998gb,misiakQED,Chetyrkin:1997vx}. 
On the other hand, a change of scheme  can be in general accommodated
by an additional non--linear term in the transformation \Eq{transf}
\beq \label{transfb}
\vec{C}(\mu) =  \left ( 1 + \f{\alpha_s (\mu)}{4\pi} \Delta
  \hat{r}_s^T + \f{\alpha}{4\pi} \Delta \hat{r}_e^T \right )
\hat{{\cal R}}^T \vec{C}^\prime(\mu) \, , 
\eeq
where $\Delta \hat{r}_s$ and $\Delta \hat{r}_e$ are matrices that
depend on the way the projection on the space of physical operators is
implemented in the effective theory calculation (which in turn
corresponds to a definition of evanescent operators).  At the order we
are interested in, they affect only $C_1$ and $C_2$. 

In the following, for definiteness,  we follow the convention of
\cite{Chetyrkin:1998gb,Chetyrkin:1997vx},  whose basis of physical
operators is a subset of \Eq{operatorbasis}.  Recalling that
$C_{2,e}^{(1)}(\mw)$ was obtained in \cite{previous} in the {\it
  standard} basis, we have calculated the matrix  $\Delta \hat{r}_e$
that connects  the two different schemes.  As a result, the
non--vanishing $\order(\aem)$ contributions to the Wilson coefficients
of the four quark operators and of the evanescent operators at
$\mu=\mw$ are given by   
\beq \label{Ce1}
\begin{split}
C_{2,  e}^{(1)} (\mw) & = -\f{22}{9} + \f{4}{3} \ln
\f{\mz^2}{\mw^2} + \frac19\, , \\
C_{3,  e}^{(1)} (\mw) & = -\f{1}{\SWS} \left ( \f{4}{9} B_0 (x_t)
  + \f{2}{9} C_0 (x_t) \right ) \, , \\
C_{5,  e}^{(1)} (\mw) & = \f{1}{\SWS} \left ( \f{1}{9} B_0 (x_t) +
  \f{1}{18} C_0 (x_t) \right ) \, , \\
C_{7,  e}^{(1)} (\mw) & = 4 C_0 (x_t) + \widetilde{D}_0 (x_t)
-\f{1}{\SWS} \left ( \f{10}{3} B_0 (x_t) - \f{4}{3} C_0 (x_t) \right )
\, , \\ 
C_{9,  e}^{(1)} (\mw) & = \f{1}{\SWS} \left ( \f{5}{6} B_0 (x_t) -
  \f{1}{3} C_0 (x_t) \right ) \, , \\
C_{1, e}^{E (1)} (\mw) & = \f{1}{\SWS} \left ( \f{4}{3} B_0 (x_t) +
  \f{2}{3} C_0 (x_t) \right ) \, , \\ 
C_{2, e}^{E (1)} (\mw) & = 4 C_0 (x_t) + \widetilde{D}_0 (x_t) +
\f{1}{\SWS} \bigg (10 B_0 (x_t) - 4 C_0 (x_t) \bigg ) \, , \\
C_{3, e}^{E (1)} (\mw) & = 4 C_0 (x_t) + \widetilde{D}_0 (x_t) \, ,
\end{split}
\eeq
with 
\beq
\begin{split}
B_0 (x_t) &= -\f{x_t}{4 (x_t - 1)} + \f{x_t}{4 (x_t -
  1)^2} \ln x_t \, , \\ 
C_0 (x_t) &= \f{x_t ( x_t-6)}{8 (x_t - 1)} + \f{x_t (2 + 3
  x_t)}{8 (x_t - 1)^2} \ln x_t \, , \\ 
\widetilde{D}_0 (x_t) &= \f{16 - 48 x_t + 73 x_t^2 - 35
  x_t^3}{36 (x_t - 1)^3} + \f{-8 + 32 x_t - 54 x_t^2 + 30 x_t^3 - 3
  x_t^4}{18 (x_t - 1)^4} \ln x_t \, .   
\end{split}
\eeq
Here we have left explicit the extra scheme dependent term $1/9$ in $C_{2,
  e}^{(1)}(\mw)$: it is numerically very small.

\begin{figure}[t]
\vspace{-7cm}
\begin{center}
\begin{tabular}{cc}
\hspace{-2.5cm}\mbox{\epsfxsize=12.5cm\epsffile{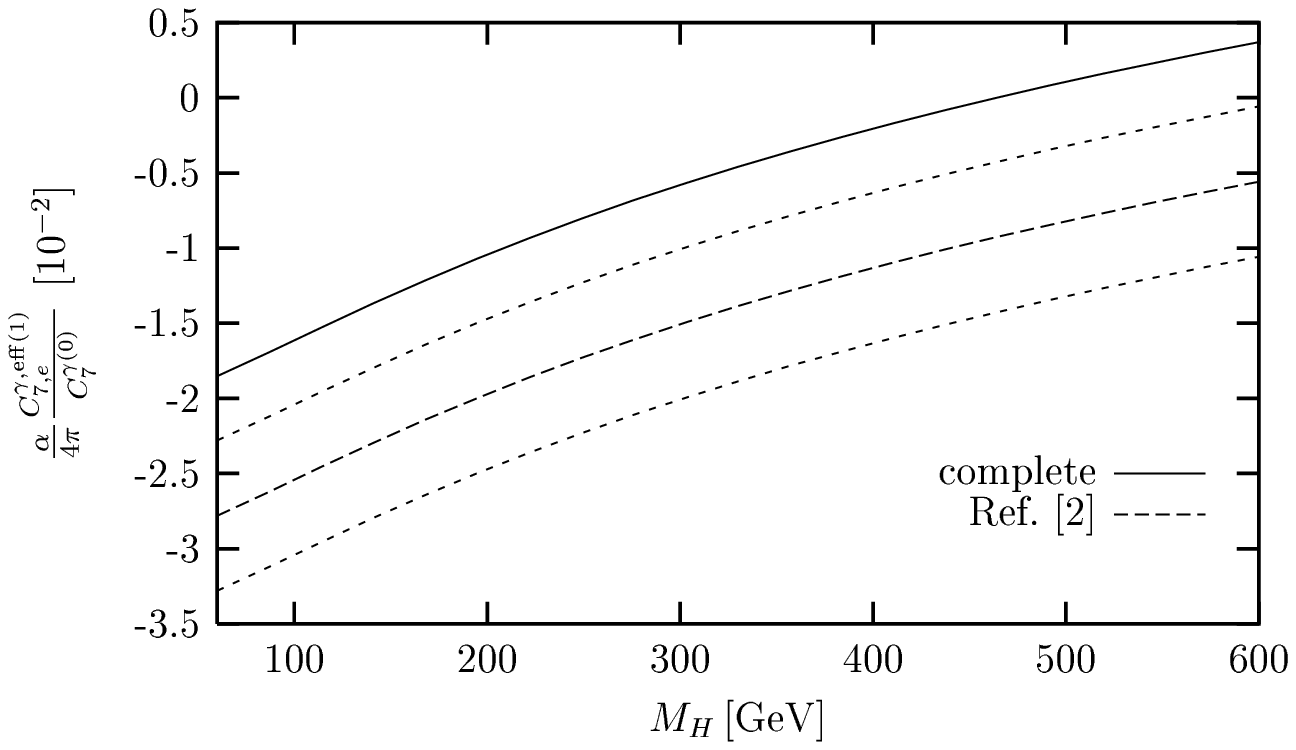}}&
\hspace{-4.5cm}\mbox{\epsfxsize=12.5cm\epsffile{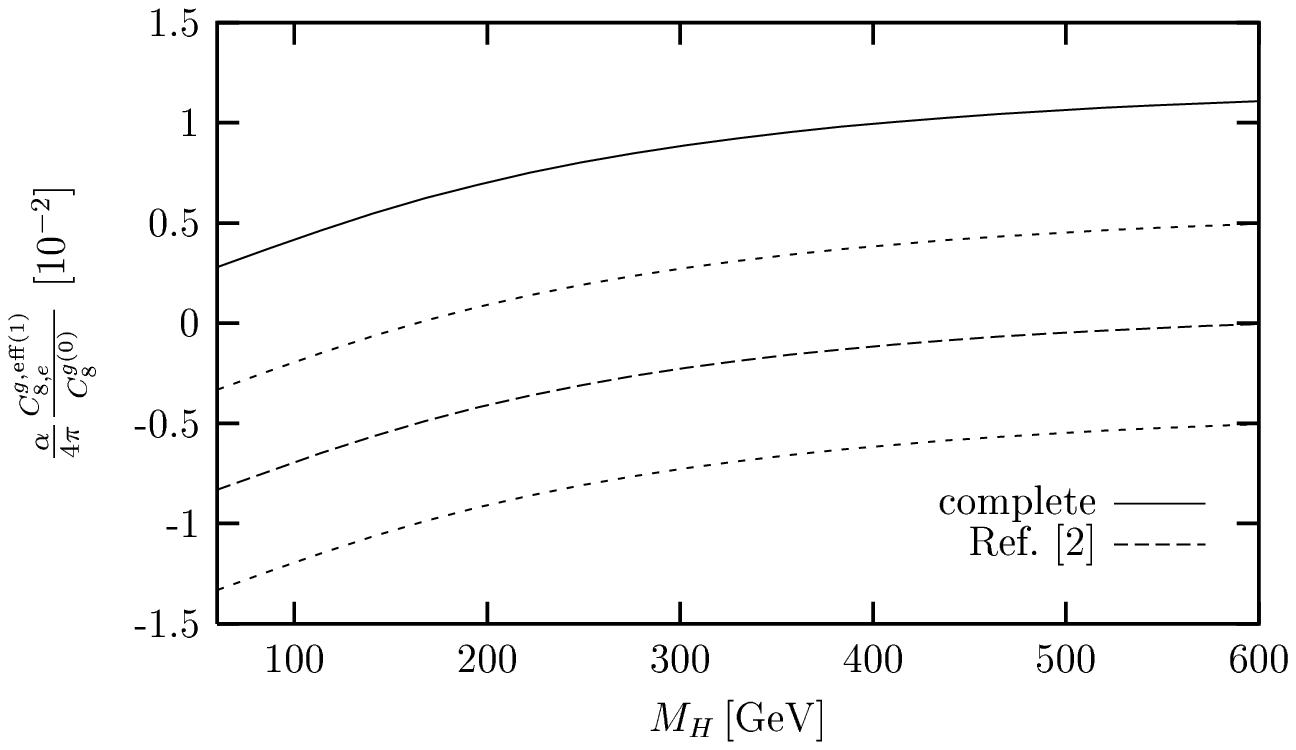}}
\end{tabular}
\end{center}
\vspace{-6.25cm}
\caption{\sf Electroweak corrections to the Wilson coefficients
$C_{7}^{\gamma}(\mw)$ and $C_{8}^{g}(\mw)$. The dashed lines represent
the results of \cite{previous} with their error estimates, the solid
lines the complete corrections to the Wilson coefficients at
$\mw$.\label{fig1}} 
\end{figure}

The final results for $C_{7, e}^{\gamma (1)}$ and $C_{8, e}^{g (1)}$
are quite  lengthy. We give  instead two accurate approximate formulas
for the $\order(g^2)$ contributions to $C_{7}^{\gamma, {\rm
    eff}}(\mw)$ and $C_{8}^{g, {\rm eff}}(\mw)$, which are valid when
the effective Hamiltonian is normalized in terms of $G_F$ as in
\Eq{hamiltonian}:   
\bea \label{fits}
\begin{split}
C_{7, e}^{\gamma, {\rm eff}(1)} (\muw) & = \f{1}{s_\smallW^2}
\left[ 1.11 - 1.15 \left ( 1 - \f{\mtbar^2}{170^2} \right ) - 0.444
  \ln \f{\mh}{100} - 0.21 \ln^2 \f{\mh}{100} \right.\\ 
& \hspace{2cm} \left. - 0.513 \ln \f{\mh}{100} \, \ln \f{\mtbar}{170}\right ] 
  +  \left( \f{8}{9} C_{7}^{\gamma (0)} - \frac{104}{243} \right)
  \ln \f{\muw^2}{\mw^2} \, , \\ 
C_{8, e}^{g, {\rm eff}(1)} (\muw) & =\f{1}{s_\smallW^2}
\left[-0.143 + 0.156 \left ( 1 - \f{\mtbar^2}{170^2} \right ) - 0.129 \ln
  \f{\mh}{100} - 0.0244 \ln^2 \f{\mh}{100} \right.\\ 
& \hspace{2cm} \left. - 0.037 \ln \f{\mh}{100} \, \ln \f{\mtbar}{170}\right ] 
  +  \left ( \f{4}{9} C_{8}^{g (0)} - \f{4}{3} C_{7}^{\gamma
      (0)} -\f{58}{81} \right ) \ln \f{\muw^2}{\mw^2}  \, .
\end{split}
\eea
Here $\mh$ is the Higgs boson mass expressed, like $\mtbar$ in GeV.  
In \Eq{Cexpansion} we use the coupling $\aem(\muw)\approx 1/128$,
while in general we employ $s^2_\smallW=0.23$, corresponding to
$g^2=4\sqrt{2} G_F \mw^2$, $\mw=80.45 \gev$ and  $\mz=91.1875
\gev$. 
\Eqs{fits} reproduce accurately (within 1.5\%) the analytic results 
in the ranges $80 \gev < \mh <300 \gev$ and $160 \gev < \mtbar <180 \gev$.
We stress that \Eqs{fits} are independent of the choice of the scale
$\mu_t$ in the QCD top mass definition: it is sufficient to calculate
$\mtbar(\mu_t)$ and employ it in \Eqs{fits}. Different choices of
$\mut$ lead to different NLO QCD corrections, but they are higher
order effects as far as the present calculation is concerned. The
$\muw$ dependence of the effective coefficients agrees with
\cite{marciano,kagan,misiakQED}.

\newpage
The size of the \ew corrections to $C_{7} ^{\gamma, {\rm eff}}$ and 
$C_{8}^{g, {\rm eff}}$  relative to the one--loop results is shown in 
Fig.~\ref{fig1} as a function of the Higgs mass. 
To compare directly the results in \Eq{fits} with the approximate
ones in \cite{previous}, we have used the same central value
$\mtbar=175.5 \gev$ in the plots. First, notice that the Higgs mass
dependence is identical, as should be expected since all the diagrams
involving the Higgs boson also involve a charged boson. Therefore, these
diagrams are not sensitive to the $Z$--$W$ mass difference or to
$\order (s_\smallW^2)$ couplings. Numerically, we see from
Fig.~\ref{fig1} that the difference is  larger than estimated in 
\cite{previous}. Although an expansion of the results in powers of
$s^2_\smallW$ converges quickly, it turns out that its second term of
$\order(\SWS)$  is larger than naively expected, and that the
two--loop correction is very sensitive to the $\mz$--$\mw$
difference. 

\section{QED--QCD evolution and the decay $B\to X_s \gamma$}

The relevant quantity in the evaluation of BR$_\gamma$ is the
effective Wilson coefficient of the magnetic operator at a scale
$\mu_b\approx m_b$. In the  resummation of  QED and QCD logarithms  
one usually keeps only terms linear in the electromagnetic coupling
$\aem$, whose running is also neglected. 
 In this case, the general structure of the evolution at
$O(\aem\alpha_s^n L^n)$ is well--known, see for instance
\cite{lectures,NNLO}. Using the notation $\vec C^T = (C_1, \ldots,
C_{10}, C_7^{\gamma, \rm eff}, C_8^{g, \rm eff})$ and restricting to
the physical
(on--shell) operators in \Eq{operatorbasis}, the coefficients at a
scale $\mu$ are given in terms of the coefficients at the scale $\mw$
by 
\begin{equation}\label{cmu}
\vec C (\mu) = \vec C^{(0)} (\mu)+ \f{\as(\mu)}{4\pi} \vec C^{(1)}_s (\mu)
+\f{\aem}{4\pi} \vec C_{e}^{(1)} (\mu) = \hat U(\mu,\mw,\aem) \,\vec
C(\mw) \, ,  
\end{equation}
where
\begin{equation}\label{utot}
\hat U(\mu,\mw,\aem)=\hat U^{(0)}(\mu, \mw)+
\hat U^{(1)}(\mu, \mw)+ \frac{\aem}{4\pi}
\left[\hat R^{(0)} (\mu, \mw) +\hat R^{(1)} (\mu, \mw)  \right] \, . 
\end{equation}
The first two terms give the pure QCD evolution. The matrices
$\hat U^{(i)}$ and $\hat R^{(i)}$ are determined by  the anomalous
dimension matrix of the operators in question and by the QCD $\beta$
function. Explicit expressions for $\hat U^{(0)}$, $\hat U^{(1)}$, and
$\hat R^{(0)}$ can be extracted from
\cite{Chetyrkin:1998fm,misiakQED}. $\hat R^{(1)}$ is presently
unknown: it requires the evaluation of the two and three--loop 
anomalous dimension matrix at $\order(\aem\as)$. From the point of
view of the expansion in $\as$ in the renormalization group improved
perturbation theory, $\hat U^{(0)}$ and $\hat U^{(1)}$  are
$\order(1)$ and $\order(\as)$, respectively. $\hat R^{(0)}$ and $\hat
R^{(1)}$  are $\order(1/\as)$ and $\order(1)$, respectively. Expanding
in $\as$ and $\aem$ we obtain 
\begin{align} \label{cfin}
\vec C (\mu) & = \hat U^{(0)}(\mu,\mw) \left[\vec
  C^{(0)} +\frac{\as(\mw)}{4\pi} \vec C_s^{(1)} \right] + \hat
U^{(1)}(\mu,\mw) \vec C^{(0)} \\ 
& + \frac{\aem}{4\pi}\left[ \hat U^{(0)}(\mu,\mw) \vec C_{e}^{(1)} +
  \hat R^{(0)}(\mu,\mw) \left(\vec C^{(0)}+\frac{\as(\mw)}{4\pi} \vec
    C_s^{(1)} \right)+ \hat R^{(1)}(\mu,\mw) \vec C^{(0)} \right] \,
. \non 
\end{align}
All the coefficients on the right hand side are understood at the
scale $\mw$. The first line results from pure QCD evolution. The 
second line mixes QED, electroweak, and QCD effects. After the
calculation of the two missing elements of $\vec C_{e}^{(1)}$
in Section 2,  the only unknown part of \Eq{cfin} relevant for
(chromo)magnetic decays is the last term. 

Turning to the particular case of $C_7^{\gamma,{\rm eff}}$ and neglecting the 
unknown term $\hat R^{(1)}(\mu,\mw) \vec C^{(0)}$, 
we see from \Eq{cfin} and \cite{lectures,misiakQED} that the
$\order (\aem)$ terms are given by   
\beq\label{UR}
C_{7, e}^{\gamma, {\rm eff}(1)}(\mu_b) = C_{7, e}^{\gamma,U}(\mu_b)
+ C_{7, e}^{\gamma,R}(\mu_b) \, . 
\eeq
The first term on the right hand side corresponds to  $\hat
U^{(0)}(\mu,\mw) \vec C_{e}^{(1)}$ in \Eq{cfin} and takes the form
\beq
\begin{split}
C_{7, e}^{\gamma,U}(\mu_b) & = \eta^{\f{16}{23}} C_{7, e}^{\gamma,
  {\rm eff}(1)}(\mw) + \f{8}{3}
\left(\eta^{\f{14}{23}}-\eta^{\f{16}{23}}\right) C_{8, e}^{g,{\rm
    eff}(1)}(\mw) \\    
& -(0.448- 0.49\,\eta) \,C_{2, e}^{(1)}(\mw) +
(0.362- 0.454 \,\eta)\, C_{3, e}^{(1)}(\mw)\\
& + (5.57- 5.86\,\eta)\,C_{5, e}^{(1)}(\mw)
+(0.321- 0.47\,\eta) \,C_{7, e}^{(1)}(\mw)\\
&+(1.588- 2.89\,\eta)\, C_{9, e}^{(1)}(\mw) \, ,
\end{split}
\eeq
where $\eta= \as(\mw)/\as(\mu_b)\approx 0.56$ for $\mu_b=m_b$. 
Here we have given analytically only
the cofactors of $C_{7}^{\gamma,{\rm eff}}$ and  $C_{8}^{g,{\rm
    eff}}$. The other terms are more involved and are given
in an approximate form, valid within 1\% 
for values of $\eta$ between 0.5 and 0.6.
However, they can all be
easily determined from the anomalous dimension matrices given in
\cite{misiakQED}\footnote{Table 2 in \cite{previous} 
  allows to change from the basis of \cite{misiakQED} to that of
  \cite{lectures}.}. The last four terms have been given in 
\cite{previous} in the operator basis of \cite{lectures}.

The second term in \Eq{UR} corresponds to 
$\hat R^{(0)}(\mu,\mw)\big(\vec C^{(0)} + \as(\mw)/(4\pi) \vec
  C_s^{(1)} \big)$ in \Eq{cfin} and is given by
\bea
\label{ew}
& C_{7, e}^{\gamma, R}(\mu_b) = 
 \f{4\pi}{\as(\mu_b)}\Bigg[ 
\left( \f{88}{575} \eta^{\f{16}{23}}
  -\f{40}{69} \eta^{-\f{7}{23}}+ \f{32}{75} \eta^{-\f{9}{23}}\right)
\left(C_7^{\gamma,{\rm eff}(0)}(\mw)+ \f{\as(\mw)}{4\pi}
  C_{7,s}^{\gamma,{\rm eff}(1)}(\mw)\right) \non\\
& \ \ \ \ \ + \left( \f{640}{1449} \eta^{\f{14}{23}}-\f{704}{1725}
  \eta^{\f{16}{23}} + \f{32}{1449} \eta^{-\f{7}{23}} -\f{32}{575}
  \eta^{-\f{9}{23}} \right) \left(C_8^{g,{\rm eff}(0)}(\mw)+
  \f{\as(\mw)}{4\pi} C_{8,s}^{g,{\rm eff}(1)}(\mw)\right) \non \\
& \hspace{-3.25cm} - 0.0449 + 0.2504 \eta - 0.236 \eta^2 \Bigg ] + (0.15
-0.178 \,\eta) \,\eta \, C_{1,s}^{(1)}(\mw) \non \\
& \hspace{-8.50cm} - (0.381 - 0.556 \,\eta)\, \eta \,C_{4,s}^{(1)}(\mw)
\, .  
\eea
The $\order(\as)$ coefficients $C_{i,s}^{(1)}(\mw)$ can be found in  
\cite{Chetyrkin:1997vx,Ciuchini:1998xe,Buras:1998xf}. The
approximate expressions are valid within 1\% for $0.5<\eta<0.6$.
 
We are now ready to give a numerical value for the $\order (\aem)$
Wilson coefficient at $\mu_b=4.7 \gev$ using \Eq{ew}.
Renormalizing the top mass at $\mu_t=\mtbar=165 \gev$, we find  
\beq\label{c7emb}
C_{7, e}^{\gamma, {\rm eff}(1)}(\mu_b)=
4.172 - 1.312  \ln \f{\mh}{100} - 0.615 \ln^2 \f{\mh}{100}
 + 2.360 \, ,
\eeq
where the first three terms correspond to $C_{7,e}^{\gamma,U}$ 
and the last one to $C_{7,e}^{\gamma,R}$. Notice that for a light
Higgs boson the first term (formally $\order(\aem \alpha_s^n L^n)$) is
twice the second one (formally $\order(\aem \alpha_s^{n-1} L^n)$). We
interpret this as evidence that purely \ew $\order(\aem_\smallW)$
effects are dominant with respect to purely QED effects.

To see how \ew corrections affect the calculation of BR$_\gamma$ it is
sufficient  to recall that, for $\mu_b = \mb$, the perturbative QCD
expression for the $b\to s \gamma (g)$ decay is proportional to 
\beq\label{br}
\left|C_7^{\gamma,\rm eff}(\mb)+ \vec{C}^{(0)} (\mb) \cdot \left (
\f{\alpha_s(\mb)}{4\pi} \vec{r}_s + \f{\aem}{4\pi} \vec{r}_e \right )
\right|^2 +B(E_0) \, , 
\eeq
where $E_0$ is the maximal photon energy in  the $b$--quark frame and
$B(E_0)$ originates from bremsstrahlung diagrams. $\vec{r}_s$ and
$\vec{r}_e$ originate from the $\order(\as)$ and $\order(\aem)$ matrix
elements of the physical operators. $\vec{r}_s$ has been computed in
\cite{matrix} with the exception of $r_{3, s}, \ldots, r_{10, s}$. It is
easy to see from these papers that\footnote{Because of the definition
  of $B(E_0)$ adopted in \cite{gambino-misiak}, we use here $r_{7
    \gamma, s} = \f{8}{9} (4 - \pi^2)$ as given in \cite{matrix}. This
  convention is different from the one of \cite{Chetyrkin:1997vx}.}  
\beq
r_{1, e} = -\f{2}{9} r_{2, s} \, , \hspace{1cm} 
r_{2, e} = -\f{1}{6} r_{2, s} \, , \hspace{1cm} 
r_{7 \gamma, e} = \f{1}{12} r_{7 \gamma, s} - \f{1}{4} r_{8 g, s} \, , 
\eeq  
while $r_{8 g, e} = 0$. Numerically, the effect of $r_{2, s}$ and  
$r_{7 \gamma, s}$ in the calculation of the inclusive branching ratio 
is quite important. On the other hand, the $\order(\as)$ contribution 
to $B(E_0)$ changes BR$_\gamma$ by less than 4\% if $1 \gev < E_0 < 2
\gev$. We therefore conclude that the only potentially relevant QED
matrix elements are the virtual corrections parameterized by $r_{1,
  e}$, $r_{2,e}$ and $r_{7\gamma,e}$. Hence, in our numerics we will
neglect the unknown last term in \Eq{cfin}, the unknown QCD matrix
elements (which are in any case suppressed by  small Wilson
coefficients), and the remaining real and virtual QED contributions to
the matrix elements. It has been observed in \cite{gambino-misiak}
that splitting the charm and top quark contributions in \Eq{br} and
normalizing them in an asymmetric way leads to an improved
perturbative QCD expansion. In the evaluation of the \ew corrections,
however, this would be an unnecessary complication.

We stress that, as we  neglect the last term in \Eq{cfin} and
some contributions to the matrix elements, our evaluation of 
$\order (\aem\alpha_s^n L^n)$ effects in $B\to X_s \gamma$ is
incomplete, although we are confident that it should provide a good
approximation.
Our numerical result is valid
in the NDR scheme supplemented by the definition of evanescent
operators  of \cite{Chetyrkin:1998gb,Chetyrkin:1997vx}. 
An analysis of the way the scheme dependent terms recombine can be
found in \cite{NNLO}. The scheme dependence of our result
is introduced in \Eq{Ce1} and in the $r_{i,e}$. 
The one from \Eq{Ce1} is numerically negligible (less than
0.01\% on BR$_\gamma$). All the residual scheme dependent pieces
would be cancelled by corresponding terms in the anomalous dimension
matrix at $\order (\aem \as)$, if it were available.  

Additional $\order(\alpha)$ contributions are introduced by
normalizing BR$_\gamma$ in terms of the semileptonic branching ratio, as
there are well--known QED corrections to the semileptonic decay
amplitude \cite{sirlin}. Unfortunately, only the leading logarithmic
term is known. The final $\order(\alpha)$ contribution to the 
expression under absolute value in \Eq{br} is therefore \cite{gambino-misiak}
\beq
\begin{split}
\varepsilon_{\rm ew}= 
\f{\aem}{4\pi} \Big [&C_{7,e}^{\gamma,\rm eff(1)} (\mb) + r_{1, e}
C_1^{(0)}(\mb) + r_{2, e} C_2^{(0)}(\mb) \\  
  &+ r_{7 \gamma, e} C_{7}^{\gamma,{\rm eff}(0)}(\mb) - 4
  \,C_7^{\gamma,\rm eff(0)}(\mb) \ln \f{\mz}{\mb} \Big] \, .  
\end{split}
\eeq
Using the reference value $\mh=115 \gev$ and $m_c/\mb=0.22$ as in 
\cite{gambino-misiak}, we find numerically
\beq
\varepsilon_{\rm ew}= 0.0025 + 0.0014 + 0.0004 + 0.0028 = 0.0071 \, , 
\eeq
which updates Eq.~(4.6) of \cite{gambino-misiak}. Here the first and
second terms correspond to the $U$ and $R$ components of
$C_{7,e}^{\gamma,\rm eff(1)}(\mu_b)$ (in
\cite{gambino-misiak} they were 0.0035 and 0.0012, respectively). 
The third term
derives from the QED matrix elements and was not included in the paper
mentioned above.  The last term, 0.0028, is due to the QED corrections
to the semileptonic decay amplitude and is the same as in
\cite{gambino-misiak}.  Notice that the first term, although
formally suppressed with respect to the second one, is larger, as it
incorporates all purely \ew contributions. The total effect of the QED
and \ew corrections in $\varepsilon_{\rm ew}$ on the branching ratio
is a 3.6\% reduction while the $\order(\aem \alpha_s^n L^n)$
contributions alone lead to a 1.6\% reduction. As different contributions
accidentally compensate each other, $\varepsilon_{\rm ew}$ is almost
exactly the same that was used in \cite{gambino-misiak}. Incorporating all 
perturbative and non--perturbative QCD corrections and using the same
numerical inputs as in  \cite{gambino-misiak}, we therefore obtain for
different values of the cutoff photon energy in the $\bar{B}$ meson
frame the following results   
\bea
\begin{split}
{\rm BR}\left[\bar{B} \to X_s \gamma\right]_{E_{\gamma} > m_b/20} 
& = (3.74\pm 0.30) \times 10^{-4} \, , \\
{\rm{BR}}\left[\bar{B} \to X_s \gamma\right]_{E_{\gamma} > 1.6 \gev} 
& = (3.61 \pm 0.30)\times 10^{-4} \, ,  
\end{split}
\label{res}
\eea
which are very close to those given in the paper mentioned above. Here the 
errors are estimates of theoretical errors also based on the analysis of  
\cite{gambino-misiak}. One can compare the first of these two results
with the present experimental world average ${\rm BR}_\gamma = (3.23 
\pm 0.42) \times 10^{-4}$ \cite{exp}.     

\section{Conclusions}
We have calculated the complete $\order (\aem)$ Wilson coefficients
relevant for radiative weak decays and described the implementation of
$\order (\aem\alpha_s^n L^n)$ effects in detail, including also the
dominant QED matrix elements.  The final impact of these
contributions on the  branching ratio of $B\to X_s\gamma$ is roughly
1.6\% for a light Higgs mass $\mh\approx 100 \gev$, and decreases
slowly for larger values of $\mh$. 

We have discussed in detail the role played by unphysical operators
in the calculation. We have adopted two different
methods to regulate the IR divergences and clarified the subtleties
that arise in the two cases. In contrast to the off--shell $\order(\as)$
calculation \cite{Ciuchini:1998xe,Bobeth:1999mk}, evanescent operators
turn out to play a crucial role in the $\order(\aem)$ computation. We
have also explained the relevance 
of gauge variant operators in our calculation. 

Our results improve
upon existent calculations \cite{previous,misiakQED,kagan,marciano}
and put \ew corrections to $B\to X_s \gamma$ on a firmer basis,
although numerically the change is negligible. The dependence of
$C_7^{\gamma, \rm eff}(\mu_b)$ on heavy degrees of freedom is now
completely known at $\order(\alpha)$. We have also included the
dominant $\order(\aem)$ matrix elements.  Still, not all the $\order
(\aem\alpha_s^n L^n)$ contributions to radiative decays  are under
control. The uncalculated corrections are related to the QED--QCD
evolution (last term in \Eq{cfin}) and to some suppressed QED
matrix elements. As the \ew contributions to QCD and \ew penguin
operators are relevant to our discussion, some two--loop QCD matrix
elements are also still missing. The incompleteness of our calculation makes it
scheme--dependent, but, as we have noted above, the scheme dependence
is remarkably small. On the other hand, the calculation of the missing
contributions would require a significant effort. In the meanwhile,
we note that: (i) the leading $\order (\aem\alpha_s^{n-1}
L^n)$ corrections affect the branching ratio of $B\to X_s \gamma$ only
in a minor way ($-0.6\%$) and (ii) the QED matrix elements are very
small,  although formally of the same order of the matching
corrections.  Therefore, one might expect the missing subleading QED
effects to be eventually small.   

\vspace{1cm}

We are grateful to Antonio Grassi and Miko{\l}aj Misiak for many helpful
discussions and to Andrzej Buras for a careful reading of the manuscript.

\end{document}